\documentclass[twocolumn,showpacs,preprintnumbers,amsmath,amssymb]{revtex4}
\usepackage{graphicx}   
\usepackage{dcolumn}    
\usepackage{bm}         

\begin{document}

\title{Localization of DNA damage by current exchanging repair enzymes: \\
effects of cooperativity on detection time} 

\author{Kasper Astrup Eriksen\\
}
\email{kasper@thep.lu.se}
\affiliation{
Department of Theoretical Physics, Lund University,\\
S{\"o}lvegatan 14A, SE-223 62 Lund, Sweden.}

\date{\today
}

\renewcommand{\thefootnote}{\arabic{footnote}} 
\setcounter{footnote}{0}

\begin{abstract}
How DNA repair enzymes find the relatively rare sites of damage is not known in great detail. Recent experiments and molecular data suggest that the individual repair enzymes do {\em not} work independently of each other, but rather interact with each other through currents exchanged along DNA. A damaged site in DNA hinders this exchange and this makes it possible to quickly free up resources from error free stretches of DNA. Here the size of the speedup gained from this current exchange mechanism is calculated and the characteristic length and time scales are identified. In particular for Escherichia coli we estimate the speedup to be $50000/N$, where $N$ is the number of repair enzymes participating in the current exchange mechanism. Even though $N$ is not exactly known a speedup of order 10 is not entirely unreasonable. Furthermore upon over expression of repair enzymes the detection time only varies as $N^{-1/2}$ and not as $1/N$. This behavior is of interest in assessing the impact of stress full and radioactive environments on individual cell mutation rates.   
\end{abstract}

\maketitle

Functional and properly working DNA is a prerequisite for the survival of all cells and organisms. However all the time localized  errors in the DNA string are introduced both by external stress factors and normal cellular functions like metabolism. It is thus of utmost importance to quickly detect the damage and repair the DNA. In Escherichia Coli an enzyme MutY is responsible for localizing 'oxidative damage' to single base pairs. Homologous enzymes presumably with a similar mode of action is found in most other species.  It is well known that MutY is able to slowly move along the 
DNA \cite{MutY_progression_Biochemistry42(2003)} while it checks the integrity of each base pair. However it has previously been pointed out that MutY acts surprisingly fast \cite{Newscientist}. In fact it probably localizes a defect base pair faster than is possible by the slow scanning above. 
Based on microscopic data it has recently been hypothesized that two MutY complexes bound to DNA are able to communicate via currents in the DNA and in this way speedup the localization process \cite{barton_PNAS_2003}. More specifically an error free stretch of DNA is a good conductor, while a defect base pair introduces a huge resistance
\cite{delaney_joc_2003}; thus if a MutY enzyme receives an electron from an upstream MutY enzyme it knows the stretch of DNA ahead of it is error free and it then detach from the DNA and instead find another stretch of DNA to scan. Intuitively it is clear this quick freeing up of MutY enzymes from error free stretches of DNA speeds up the localization of damaged base pairs, but by how much?
There are two relevant time scales in the proposed process. The first is the time $T$ it on average takes to localize a damaged base pair by slowly scanning the DNA, without utilizing the currents. The second is the time $\tau$ it takes to realize a stretch of DNA is without errors using the current, i.e.\ $ \tau$ is the time from a MutY enzyme attach to an error free piece of DNA until it detaches and goes somewhere else to attach. In this paper, we first show that the time it takes for MutY in the above scheme to localize a damaged base pair is roughly $\sqrt{T \tau}$ corresponding to a speedup of order $\sqrt{T/\tau}$. This expression for the speedup is even retained  in the presence of many other kinds of current interacting repair enzymes. However $T$ is in this case the time until the first repair enzyme, which need not be MutY, locates the error by scanning alone.

\section{Model}
\begin{figure}[htbp]
  \centering
  \includegraphics*[width=7cm]{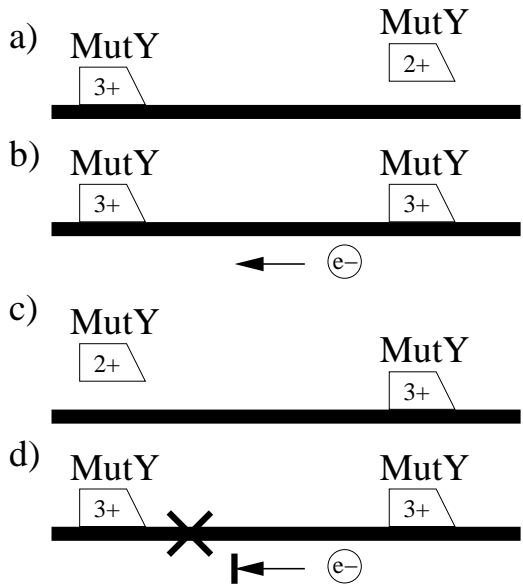}
  \caption{
a) The left MutY repair enzyme is bound to DNA and slowly progress to the right while it scans the integrity of the base pairing. The [4Fe4S] cluster in MutY is in a +3 charge configuration when bound to DNA, while it is in a +2 configuration when not bound (right MutY). 
b) Upon binding to DNA the right MutY enzyme emit an electron onto DNA and changes the charge of its [4Fe4S] cluster from +2 to +3.
c) If the DNA is free of errors the emitted electron travels along the DNA until it reaches the left MutY enzyme. Here the electron changes the charge of the [4Fe4S] cluster to +2 and thus destabilizes the DNA binding of the left MutY enzyme. The left MutY enzyme  then attach to and scan a different section of DNA more likely to posses an error.
d) If on the other hand the DNA piece in between the two MutY enzymes contains an error the electron never reaches the left MutY enzyme, which then keep scanning the DNA until it reaches and fixes the error. The current exchange thus selectively free up resources from error free patches of DNA.
}
\label{Fig1}
\end{figure}

The model is pictorially presented in Figure \ref{Fig1}. 
See also \cite{Newscientist,barton_PNAS_2003}.
The repair enzyme MutY contains an evolutionary well-conserved [4Fe4S] cluster that is suspected to change its charge configuration from +2 to +3 upon binding to DNA \cite{barton_PNAS_2003}. Upon binding to DNA an electron is thus emitted into DNA, while upon receipt of an electron from DNA the MutY-DNA binding is destabilized. As only error-free stretches of DNA are able to transport the electron from one MutY enzyme to a neighboring MutY enzyme \cite{delaney_joc_2003} this charge exchange enables MuTY to quickly free up scanning resources from error free stretches of DNA \cite{barton_PNAS_2003}
--- see Figure \ref{Fig1}.  
In the original proposal only MutY enzymes participates in the current exchange and to illustrate the line of thought we first consider this scenario. Afterward, we then extend the consideration to the case where there are many different kinds of repair enzymes each  specialized to fix a specific kind of damage. Furthermore, we also consider the effect of a finite scan length before MutY spontaneously drops off DNA.

\subsection{Only MutY and it scans forever}

After a a base pair get damaged, how long time does it take before a MutY enzyme has located the error?
In the regime where current scanning yields a considerable speedup in the location of the error, the damaged base pair typically is located by a MutY enzyme that binds to DNA next to it and then scans all the way up to and including the faulty base pair (and not by a MutY enzyme that just happened to be nearby at the time of damage). Let $t_{\textrm{detect}}$ denote the typical detection time and $v$ the scan velocity of MutY. The MutY enzyme then typically docks onto DNA within a distance of the order $v t_{\textrm{detect}}$. The rate with which a MutY enzyme randomly docks onto a specific base pair and start scanning is denoted by $k$. The probability for a MutY enzyme to land within a distance of $v t_{\textrm{detect}}$ of the error in the time interval $t_{\textrm{detect}}$ can be estimated as $k v t_{\textrm{detect}}^2$. As at time $t_{\textrm{detect}}$ a MutY enzyme typically arrives at the faulty base pair this probability is of order 1 and 
\begin{equation}
t_{\textrm{detect}} \approx \frac{1}{\sqrt{k v}}\,.
\end{equation}
A more detailed model calculation in the supplementary material yields the same result apart from a factor $1.3$.
The average docking rate $k$ can be expressed as
\begin{equation}
k=\frac{N(\textrm{MutY})}{\tau L}=\frac1{v T \tau} \,,
\end{equation}
where $\tau$ is the time between two successive binding events for a single MutY enzyme.
$N(\textrm{MutY})$ is the total number of MutY enzymes and $L$ is the total number of base pairs in DNA. $T= L/v/N(\textrm{MutY})$ is the time it takes for all the MutY enzymes to scan all the bases of DNA once. From standard Poisson statistics $T$ is also the detection time of a damaged point in the traditional scenario without any cooperation between MutY enzymes. In terms of $T$ and $\tau$ the detection time is
\begin{equation}
t_{\textrm{detect}} \approx \sqrt{T \tau}\ = T \sqrt{\frac{\tau}{T}} \,. 
\end{equation}
The cooperation thus give a speed up of order $\sqrt{\tau/T}$. Before I get down to consider numbers and orders of magnitude I want to make the model a tad more realistic.

\subsection{Full model}

The functionally central [4Fe4S] cluster is also present in other repair enzymes e.g.\  endonuclease III. It is thus very likely that other repair enzymes also are able to inject currents into DNA and participate in the electrical scanning of DNA. Furthermore all of these repair enzymes then get 'attracted' to the damaged DNA pair in exactly the same way as MutY. In the above model and calculation we thus have to replace 'MutY' by 'any repair enzyme participating in the DNA mediated charge transport' --- for short just a repair enzyme below. Likewise the calculated detection time $t_{\textrm{detect}}$ becomes the time before the first repair enzyme finds the damaged site and $T$ the average time it take for any repair enzyme to find the site without using currents. I have here implicitly assumed that both the scan velocity $v$ and the single repair enzyme attempt frequency $1/\tau$ are of the same order of magnitude for all repair enzymes i.e.\ MutY is a typical repair enzyme. Biologically the time $t_{\textrm{detect}}$ is not the most relevant one as the first repair enzyme that arrives at the damaged base pair very likely is unable to fix the damage. For instance is MutY's primary target oxidation damage resulting in G:A and 8-oxo-G:A mismatched pairs.
Instead the first MutY enzyme arrives on average as the 
$N/N(\textrm{MutY})$ repair enzyme resulting in a detection time of order $N/N(\textrm{MutY}) t_{\textrm{detect}}$. Here $N$ is the total number of repair enzymes. $N/N(\textrm{MutY})$ can also be expressed as $T/T(\textrm{MutY})$, where $T(\textrm{MutY})$ is the time it takes for a MutY enzyme to find the site by scanning alone. The MutY detection time is thus
\begin{equation}
t_{\textrm{detect}}(\textrm{MutY}) \approx T(\textrm{MutY}) \sqrt{\frac{\tau}{T}} \,,
\label{result full}
\end{equation}
again corresponding to a speedup of size $\sqrt{\tau/T}$, but this time $T$ is the time it take for any repair enzyme to locate the damage.

I have not yet considered that MutY is known spontaneously to drop off DNA after having scanned of the order 100 base pairs (bp) \cite{MutY_progression_Biochemistry42(2003)}. In order to estimate the effect of this
I am going to derive the above result Eq.~(\ref{result full}) in a slightly different manner. The MutY enzyme that eventually localizes the damage typically docks onto DNA within a distance $\Delta$ from the faulty base pair, where $\Delta$ both has to be small enough to allow scanning all the way up to the error i.e.\ $<100$bp
and also so small that it is unlikely another repair enzyme dock in front of the MutY enzyme while it scans.
The last length scale can be estimated to be of the order $\Delta=\sqrt{v/k}=v\sqrt{T \tau}$. The MutY detection time $t_{\textrm{detect}}(\textrm{MutY})$ is then determined as above by 
setting the probability for a MutY enzyme to dock within a distance $\Delta$ in the time interval $t_{\textrm{detect}}(\textrm{MutY})$ equal to 1 i.e.\ 
\begin{equation}
t_{\textrm{detect}}(\textrm{MutY}) = T(\textrm{MutY}) \frac{v \tau }{ \Delta} \,, 
\end{equation}
where $\Delta = \textrm{min}(100\textrm{ bp},v \sqrt{\tau T})$. Notice that the choice $\Delta = v \sqrt{\tau T}$ leads to Eq. (\ref{result full}). 

\subsection{Estimating order of magnitude}

I have not been able to find any direct experimental measurements of $\tau$ and $T$, so I am instead going to use a somewhat more uncertain way to estimate the size of the reduction $\frac{ \Delta}{v \tau }$. The distance $v \tau$ is the average scan length of a repair enzyme (MutY) in the presence of the current interactions. The numerator $\Delta$ is the smallest of the maximal scan length 100bp and the docking distance $\sqrt{v \tau \, v T}$. In fact, I believe these two distances are of the same order of magnitude as anything else seems inefficient, so I assume $\sqrt{v \tau \, v T} \le 100$bp, with equality as the most likely option. The distance 
$v T = L/N$ is the average distance between repair enzymes. From this I can estimate 
$v \tau$. The reduction is thus  
$\frac{ \Delta}{v \tau } \ge v T/100\textrm{ bp} = 5 \cdot 10^4/N$, where $N$ is the total number of repair enzymes with a current exchange mechanism similar to MutY and I have used that the length of E.Coli's  DNA is $5 \cdot 10^6$ base pairs. Unfortunately $N$ is unknown.
In \cite{demple_ann_rev_biochem_1994} the number of the two [4Fe4S]$^{2+}$ containing repair enzymes MutY and endonuclease III is estimated to be 30 and 400 respectively. In the same paper the number of formamidopyrimidine glycosylase (FAPy or MutM) repair enzymes is estimated to 400. FAPy does not contain the [4Fe4S]$^{2+}$ cluster. The target of FAPy is 8oxoG which is estimated to constitute 5\% of all adducts due to oxidative damage \cite{beckman_jbc_1997}. All in all it seems reasonable that the total number of repair enzymes participating in the current exchange mechanism is significantly smaller than 50000, and that a speedup of order 10 is realistic. Notice this would correspond to a typical scan length 10 times smaller than the maximal one (100 bp) due to the current exchange mechanism ($v \tau$ is of the order 10 bp).

\section{Conclusion}

In this paper we have considered some of the implications of a very interesting proposal for cooperation between repair enzymes in the localization of defects in single base pairs. First we have pointed out that the mechanism is likely to speedup the localization by a factor of order 10 compared with independent scanning by the repair enzymes. If an error is detected by the 30 MutY enzymes in say 20 minutes (it has to be considerably shorter than the replication time \cite{nghiem_pnas_1988}) this corresponds to a reduction in scanning speed from 125 bp/sec to 13 bp/sec. For comparison, the scan velocity for RNA polymerase is 50 bp/sec, while for DNA polymerase it is 1000 bp/sec. Another in principle testable prediction is that upon over expression of all the repair enzymes with say a factor 4 the detection time is only decreased by a factor 2 (both $T(\textrm{MutY})$ and $T$ are 4 times smaller), because information is not carried as efficiently along the DNA any more. Physiologically oxidative and radiative environments may result in an increased expression of repair enzymes \cite{kim_mutat_res_1996}, so the square root behavior and the coupling of the effectiveness of different kinds of repair enzymes is potentially of huge importance for the mutation rates in these kinds of stress full environments.  
Summing up, only further experimentation can finally confirm this charge transport mechanism, while we here have demonstrated that the mechanism indeed offers a great benefit for the cell. In addition, the model is a nice toy model for   protein cooperativity and one might wonder if the underlying principles behind could be of practical use in apparently unrelated engineering problems.

\section*{Acknowledgments}
Kasper Astrup Eriksen acknowledges support from both the Danish Natural Science Research Council grant number 21-03-0284 and the Bio+IT programme
under the {\O}resund Science Region and {\O}forsk.

\end{document}